\documentclass[
 reprint,nofootinbib,
 superscriptaddress,
 preprintnumbers,
 amsmath
 amssymb,
 prx,
 floatfix,
]{revtex4-2}

\DeclareMathAlphabet{\mathsfit}{T1}{\sfdefault}{\mddefault}{\sldefault}
\SetMathAlphabet{\mathsfit}{bold}{T1}{\sfdefault}{\bfdefault}{\sldefault}

\usepackage{amsmath}
\usepackage{amssymb}
\usepackage{graphicx}
\graphicspath{{fig/}}   
\usepackage{dcolumn}
\usepackage{bm}
\usepackage{makecell}
\usepackage{braket} 
\usepackage{siunitx}
\usepackage{color}
\usepackage{array}
\usepackage{xr}
\usepackage{upgreek}
\usepackage[normalem]{ulem}
\usepackage[bottom]{footmisc}
\usepackage[colorlinks]{hyperref}

\usepackage[dvipsnames]{xcolor} 
\definecolor{darkred}{rgb}{0.801, 0.0, 0.0}

\hypersetup{
    colorlinks=true,
    allcolors=darkred 
}

\usepackage[capitalize]{cleveref}
\usepackage[caption=false]{subfig}
\usepackage{multirow} 
\usepackage[version=4]{mhchem} 
\usepackage[siunitx]{circuitikz}
\usepackage{tikz}
\usepackage[normalem]{ulem}

\begin{document}
\preprint{MIT-CTP/6131}
\title{Analog quantum simulation of $\phi^4$ field theory with a superconducting transmission line}

\def\RLEaffil{Research Laboratory of Electronics, Massachusetts Institute of Technology, Cambridge, MA 02139, USA}
\def\LLaffil{Lincoln Laboratory, Massachusetts Institute of Technology, Lexington, MA 02421, USA}
\def\Physaffil{Department of Physics, Massachusetts Institute of Technology, Cambridge, MA 02139, USA}
\def\EECSaffil{Department of Electrical Engineering and Computer Science, Massachusetts Institute of Technology, Cambridge, MA 02139, USA}
\def\CTPaffil{Center for Theoretical Physics - a Leinweber Institute, Massachusetts Institute of Technology, Cambridge, MA 02139, USA}
\def\AFRLaffil{Air Force Research Laboratory, Information Directorate, Rome, New York 13441, USA}
\def\ARGaffil{Physics Division, Argonne National Laboratory, Lemont, Illinois, 60439, U.S.A}

\author{Ilan~T.~Rosen}
\thanks{These authors contributed equally to this work.}
\affiliation{\RLEaffil}

\author{Neill~C.~Warrington}
\thanks{These authors contributed equally to this work.}
\affiliation{\CTPaffil}

\author{Max~Hays}
\affiliation{\RLEaffil}

\author{Christopher~McNally}
\affiliation{\RLEaffil}
\affiliation{\EECSaffil}

\author{Joshua Lin}
\affiliation{\CTPaffil}
\affiliation{\ARGaffil}

\author{Stephen~Sorokanich III}
\affiliation{\AFRLaffil}

\author{Julian~Bender}
\email{benderj@mit.edu}
\affiliation{\CTPaffil}

\begin{abstract}
    Understanding the dynamics of interacting quantum fields remains a fundamental challenge.
    We propose a superconducting transmission line for analog quantum simulation of $(1+1)$-dimensional $\phi^4$ field theory.
    The transmission line comprises fluxonium-like circuit elements whose continuous and unbounded phase variables encode a scalar field without Hilbert-space truncation.
    Their anharmonic potentials provide strong interactions.
    Tuning the circuit parameters provides access to two distinct regimes.
    In the first, the transmission line emulates $\phi^4$ field theory with a single-well potential.
    We confirm this correspondence through lattice Monte Carlo calculations of the low-energy spectra of both the circuit
    and lattice $\phi^4$ theory. 
    This regime enables scattering experiments between counter-propagating particle wave packets.
    The second parameter set yields a double-well potential that emulates $\phi^4$ theory in the symmetry-broken phase and hosts degenerate vacua.
    Interfaces between the two vacua form topological kinks, and the transmission line can simulate kink--antikink collisions.
    We present state preparation and measurement protocols for both collision experiments, enabling spatially resolved, real-time studies of particle and soliton scattering in an interacting quantum field theory.
\end{abstract}

\maketitle

\section{Introduction}
It is challenging to compute the real-time dynamics of strongly interacting quantum field theories from first principles. This results in an incomplete theoretical understanding of a number of important systems in nuclear and particle physics such as heavy ion collisions and neutron star mergers.
Questions about entanglement generation, energy flow, and non-equilibrium behavior in these systems remain challenging to answer from first principles. 

A canonical toy model for the theories encountered in nuclear and particle physics is $\phi^4$ theory.
This is a relativistic quantum field theory of a scalar field $\phi$, which is locally continuous and unbounded, and is
encoded in the Lagrange density
\begin{equation}
\label{eq:lagrangian}
    \mathcal{L} =  \frac{1}{2}\Big[\dot{\phi}^2 - c^2(\partial_x \phi)^2 - \omega_0^2 \phi^2\Big] - \frac{\lambda}{4!} \phi^4~,
\end{equation}
where $c$ is the speed of light, $\omega_0$ is the bare Compton frequency of the particle, and $\lambda$ is the coupling \cite{Peskin:1995ev}.
We include factors of $c,\,\hbar$ in expressions throughout.
This theory hosts paradigmatic features of relativistic field theories such as particle production/annihilation, spontaneous symmetry breaking, and topological excitations. Solving the dynamics of $\phi^4$ theory would be meaningful progress towards a theoretical understanding of real-time phenomena in nuclear and particle physics. 

While properties of the weakly interacting theory (small $\lambda$) are routinely determined via perturbative techniques, the real-time dynamics of strongly coupled $\phi^4$ theory are in practice difficult to determine today by any method: classical computation, quantum computation, or experiment. 
Classically, the sign problem hinders numerical lattice calculations, while tensor network approaches suffer from the continuous and unbounded nature of the field variable~\cite{Sayegh:2025rlp,Milsted:2013rxa}. 
Present-day digital quantum computers face challenges in simulating continuous, unbounded fields because of the qubit
numbers and circuit depths required to encode the fields and implement their interacting dynamics.
Finally, while particle colliders routinely do scattering experiments, they only probe the spatially-asymptotic states that reach the detector, and only partial information about the scattering event itself is available by reconstruction.

Analog quantum simulation is a promising approach to study the real-time dynamics of quantum systems.
An analog simulator is an engineered quantum environment that emulates a target system by mimicking its Hamiltonian.
Dynamics of the target system may then be studied by allowing the simulator to time-evolve. 
Some analog simulators architecturally resemble digital quantum computers~\cite{hensgens2017, Bluvstein2022, Rosen2024, Google2025simulator}.
Other simulators have purpose-designed architectures~\cite{cappellaro2007, Houck2012, gross2017, donnelly2026}.
Waveguide-like superconducting circuits, in particular, have been used to study condensed matter~\cite{underwood2012, Fitzpatrick2017dissipative, Kim2021ssh}, and a device emulating sine-Gordon field theory has been proposed~\cite{Roy2021sinegordon, Roy2023sinegordon}.
Designing a simulator poses the challenge of accurately reproducing the effective Hamiltonian of the target system while retaining state preparation and measurement capabilities.
Specifically, the creation of an analog quantum simulator for scalar field theories is confounded by the lack of unbounded field variables in qubit-based devices and the difficulty of realizing sufficiently strong interactions between photons in waveguides and cavities~\cite{Barrett2013, Sundaresan2015, Nielsen2025}.

\begin{figure}[t]
  \centering
  \includegraphics[width=0.4\textwidth]{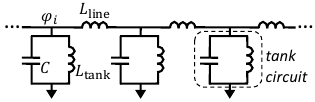}
  \caption{\textbf{The free-field transmission line} emulates a free scalar field theory. An array of tank circuits (dashed box), each consisting of an inductor $L_{\text{tank}}$ and capacitor $C$ in parallel, are inductively coupled to each other via $L_{\text{line}}$. The field variable is the node flux at each array site $\varphi_i$. The inductor that connects the ground plane to the transmission line renders the fundamental quantum field variable unbounded.}
  \label{fig:free-field-circuit-diagram}
\end{figure}

We here present an analog quantum simulation device that emulates a continuous, unbounded $\phi^4$ lattice field theory in $1+1$ dimensions. 
The device is a nonlinear quantum microwave transmission line fashioned as a superconducting circuit. 
Its architecture draws from the fluxonium qubit circuit~\cite{Manucharyan2009Fluxonium, Yan2020EngineeringFramework, Ye2021QuartonPRL} and provides exceptionally strong interactions between microwave photons, emulating particle-particle interactions in a non-perturbative regime. 
We provide parameter choices by which (1) the device accurately reproduces the static and dynamical behavior of continuum $\phi^4$ theory, and (2) the continuum limit is explicitly taken analogously to the common practice in lattice quantum chromodynamics (QCD)  \cite{Rothe:1992nt,lüscher1998advancedlatticeqcd} .
We then describe two scattering experiments that can be performed with the transmission line: a particle-particle scattering experiment enabled by deterministic preparation of particle wave packets, and scattering between kink and antikink defects within the degenerate vacuum of the field theory in the double-well potential regime.
For both, the passive dynamics of the transmission line directly emulate time evolution of the field theory, and we provide measurement protocols for local expectation values and two-point correlators.

\section{Constructing a field theory}
In this section, we construct a non-linear lumped-element microwave transmission line that encodes a $\phi^4$ quantum field theory in $1+1$ dimensions. 
The transmission line is a superconducting circuit that we describe using standard circuit quantum electrodynamics methods~\cite{Vool_2017, krantz2019quantum}.
We begin by encoding the free lattice field theory.
We then add the necessary components to realize a repulsive quartic interaction and propose parameters for the device.

\subsection{Free field theory }

The Lagrange density for a free massive scalar particle is given in Eq.  \eqref{eq:lagrangian} when $\lambda=0$.
We wish to construct a superconducting circuit that reproduces this theory in an appropriate limit.
Consider the superconducting transmission line shown in Fig.~\ref{fig:free-field-circuit-diagram}, formed by a chain of identical tank circuits composed of inductors $L_\mathrm{tank}$ and capacitors $C$, connected by in-line inductances $L_\mathrm{line}$. The Lagrangian of this circuit is $L = \sum_{i=1}^{N} L_i$, where
\begin{equation}
    L_i =  \varphi_0^2\Bigg( \frac{C}{2}   \dot{\varphi}_i^2  -\frac{1}{2L_{\text{line}}}  (\varphi_{i+1}-\varphi_i)^2 - \frac{1}{2L_{\text{tank}}}  \varphi_i^2 \Bigg),
\end{equation}
and where $\varphi_i$ is the dimensionless node flux of the $i^{\text{th}}$ node of the circuit and $\varphi_0=\hbar/2e$ is the reduced flux quantum.
Via the identification
\begin{equation}\label{eq:rule_one}
    \varphi_i \mapsto \phi_x,
\end{equation}
where $\phi_x$ is the lattice discretization of $\phi$ at position $x=\Delta x\, i$ and $\Delta x$ is the lattice spacing, this transmission line evidently encodes the free, massive scalar field theory. 

\begin{figure}[t!]
    \centering
    \includegraphics{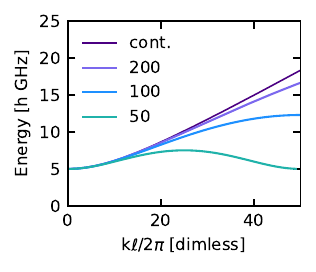}
    \caption{\textbf{Free scalar field dispersion relation.} Dispersion relations for free-field transmission lines with  parameters listed in Table \ref{tab:free-theory-circuit-params}. As the number of unit cells increases from 50 (teal), to 100 (blue), to 200 (light purple), the discrete transmission line dispersion relations approach that of the continuum theory (dark purple).}
    \label{fig:free-theory-dispersion-relation}
\end{figure}

We next wish to take the literal continuum limit of the transmission line by increasing the number of tank circuits in the device.
We will show that a faithful approximation of the continuum quantum field theory is recovered in a device of realistic scale.
To that end, here and throughout, we consider a theory with fixed total length $\ell$ but variable number of lattice sites $N$ and lattice spacing $\Delta x$ such that the product $N \Delta x = \ell$ is constant. 
Under the identification Eq.~(\ref{eq:rule_one}), the circuit Lagrangian maps to $L = \sum_x \Delta x ~ \mathcal{L}_x$, where
\begin{equation}
\mathcal{L}_x=\frac{\xi}{2} \Bigg[    \dot{\phi}_x^2  -\frac{c^2}{\Delta x^2}  (\phi_{x+\Delta x}-\phi_x)^2 - \omega_0^2 \phi_x^2 \Bigg] 
\label{eq:lag_density_line}
\end{equation}
is the Lagrange \textit{density} of the lattice field theory and
\begin{align}
    \omega_0^2 & = \frac{1}{C L_{\text{tank}}} \nonumber \\
    c^2 & = \frac{1}{C L_{\text{line}}}\frac{\ell^2}{N^2} \nonumber \\
    \xi & = \frac{N C \varphi_0^2}{\ell} \,.
\end{align}
If the circuit parameters are scaled as
$C, L_{\text{line}}  \propto N^{-1},   L_{\text{tank}}  \propto N$, then $\xi, v, \omega_0$ remain constant as $N\rightarrow \infty$, and the Lagrange density of the discrete transmission line converges to 
\begin{equation}
\label{eq:cont-lim-free-theory}
\mathcal{L}_x=\frac{\xi}{2}\Big[ \dot{\phi}_x^2  -c^2(\partial_x \phi_x)^2 -\omega_0^2 \phi_x^2\Big].
\end{equation} 
This has the same functional form as Eq. \eqref{eq:lagrangian}, so the continuum limit of the transmission line in Fig. \ref{fig:free-field-circuit-diagram} is a free massive scalar field. Note that here and throughout $c$ is to be understood as in-medium the speed of light of the transmission line.

\begin{table}[t!]
    \centering
    \renewcommand{\arraystretch}{1.3}
    \begin{tabular}{|c|c|c|c|}
    \hline
    N  & $C$ [fF] & $L_{\text{line}}$ [nH] & $L_{\text{tank}}$ [nH]  \\
    \hline
    50  & 40 & 80 & 25 \\
    100  & 20 & 40 & 50 \\
    200  & 10 & 20 & 100 \\
    \hline
    \end{tabular}
    \caption{\textbf{Circuit parameters for a free-field transmission line.} A literal continuum limit is approached by increasing the number $N$ of tank circuits in the array.}
    \label{tab:free-theory-circuit-params}
\end{table}

This claim is illustrated in Fig. \ref{fig:free-theory-dispersion-relation}, where the spectrum of a relativistic particle with rest-mass energy of $\simeq 5\, h\, \text{GHz}$ is approached by the sequence of discrete transmission lines with the parameter choices in Table \ref{tab:free-theory-circuit-params}. 
Importantly, in the frequency range of 5-\SI{10}{GHz} (a convenient operating domain for superconducting circuits), the excitation energies of the $N=200$ transmission line are virtually indistinguishable from those of the continuum theory.
Standard fabrication techniques are readily capable of manufacturing a discrete transmission line of this scale and with these parameter values~\cite{Macklin2015}.

\subsection{Interacting field theory }

We now construct a non-linear transmission line (NLTL) which emulates an {\em interacting} scalar field theory (i.e. $\lambda\neq0$). 
This device is a modification of the free-field transmission line with Josephson junctions providing non-linearity, and the strength and form of the non-linearity is adjustable through the choice of circuit parameters.

\begin{figure}[t!]
  \centering
  \includegraphics[width=0.4\textwidth]{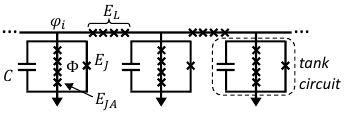}
  \caption{\textbf{The nonlinear transmission line} emulates an interacting scalar field theory. It is formed by an array of nonlinear tank circuits. Each tank circuit comprises an inductor, a Josephson junction $E_J$, and a capacitor $C$, all in parallel, and is threaded by an external magnetic flux $\Phi$. Adjacent tank circuits are inductively coupled with inductive energy $E_L$. Inductors are shown realized by superinductors; $n=4$ junctions are drawn for brevity.}
  \label{fig:interacting-field}
\end{figure}

The NLTL is shown in Fig.~\ref{fig:interacting-field}. 
As before, it comprises an array of tank circuits that are inductively coupled.
The tank circuits now adopt the fluxonium circuit, which consists of a capacitor with capacitance $C$, a single Josephson junction with Josephson energy $E_J$, and a superinductor in parallel~\cite{Manucharyan2009Fluxonium}.
The latter is a serial array of $n$ Josephson junctions with Josephson energy $E_{JA}$ and approximates a linear inductance with inductive energy $E_{JA}/n$. 
We have additionally represented the in-line inductances $L_{\text{line}}$ as superinductors for convenience; this choice of inductor technology is not essential.
Within each tank, the single Josephson junction and junction array form a loop threaded by a magnetic flux $\Phi$.
The tank circuit realizes the potential energy
\begin{equation}\label{eq:tank_potential}
   U(\varphi_i, b) =-E_J\cos(\varphi_i-b) - n E_{JA}\cos \frac{\varphi_i}{n}
\end{equation}
where we have introduced the reduced external flux $b = 2\pi\Phi/\Phi_0$ and $\Phi_0 = h/(2e)$ is the magnetic flux quantum (this is atypical notation, which we chose to avoid confusion)\footnote{Note that in the derivation of the potential energy we have assumed current conservation across the junction arrays~\cite{Kim2025Emergent}.}. 
The Lagrangian of the NLTL is $L = \sum_{i=1}^{N} L_i$, where
\begin{equation}\label{eq:lagrange_NLTL}
    L_i =  \varphi_0^2\Bigg( \frac{C}{2}   \dot{\varphi}_i^2  -\frac{1}{2L_{\text{line}}}  (\varphi_{i+1}-\varphi_i)^2 \Bigg) - U(\varphi_i, b).
\end{equation}

To map the NLTL to $\phi^4$ theory, we invoke the identification Eq.~(\ref{eq:rule_one}) and Taylor expand Eq.~(\ref{eq:lagrange_NLTL}).
We choose $b=\pi$ to enable repulsive interactions---a positive $\varphi^4$ coefficient---while suppressing odd terms.
For this choice, the coefficients of the linear ($\varphi^2$) and interacting ($\varphi^4$) terms in the potential are found by expanding about $\varphi=0$. 
One finds
\begin{equation}
\label{eq:lattice-phi4-theory}
    \mathcal{L}_x  =  \frac{\xi}{2}   \Bigg[ \dot{\phi_x}^2 - \frac{c^2}{\Delta x^2} (\phi_{x+\Delta x}-\phi_x)^2  - \omega_0^2 \phi_x^2 - \frac{2\lambda}{4!} \phi_x^4 \Bigg] ,
\end{equation}
up to corrections of $\mathcal{O}(\phi^6)$ (which we show to be small in Sec. \ref{sec:quant-non-lin}). 
The parameters of the lattice $\phi^4$ theory are related to those of the circuit through the following relations:
\begin{align}
\label{eq:int_theory_mapping}
    \omega_0^2 & = \frac{1}{\varphi_0^2 C}(E_{JA}/n-E_J) \nonumber \\
    c^2 & = \frac{1}{C L_{\text{line}}}\frac{\ell^2}{N^2} \nonumber \\
    \lambda & = \frac{1}{\varphi_0^2 C}(E_J - E_{JA}/n^3) .
\end{align}

\begin{figure}[t!]
  \label{fig:tank-potential}
  \subfloat{\label{fig:4a}}
  \subfloat{\label{fig:4b}}
  \centering
  \includegraphics{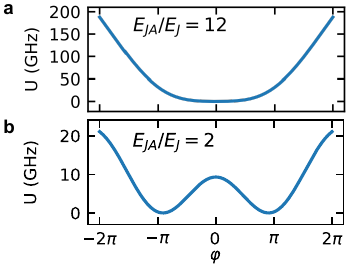}
  \caption{\textbf{Tank circuit potential.} \textbf{(a)} A heavily quartic potential well is formed when $E_{JA}/E_J>n$. \textbf{(b)} A double-well potential is formed when $E_{JA}/E_J<n$. Tank circuit parameters from Table~\ref{tab:interacting} are used.}
\end{figure}

By tuning the parameters of the circuit, different regimes of $\phi^4$ theory may be explored. 
When $E_{JA}/E_J$ slightly exceeds $n$, the tank circuit realizes a repulsive potential with strong interactions (a positive $\varphi^2$ coefficient and a large and positive $\varphi^4$ coefficient; Fig.~\ref{fig:4a}). 
When $E_{JA}/E_J<n$, the $\varphi^2$ coefficient is negative and a symmetry-breaking double-well potential is formed (Fig.~\ref{fig:4b}). 
Other potentials, including cubic nonlinearities and the quarton regime~\cite{Yan2020EngineeringFramework,Ye2021QuartonPRL}, are accessible via other parameter selections. 
Circuit parameters for realizing the strongly repulsive and double-well regimes are summarized in Table~\ref{tab:interacting}, where we have switched to energetic units $E_C=e^2/2C$ and $E_L = \varphi_0^2/L_{\text{line}}$.
For the single-well regime, we select $E_J$ and $E_{JA}$ so that the effective inductance of the tank circuit (determined by the total $\varphi^2$ coefficient) is equivalent to that of the free-field transmission line (Table~\ref{tab:free-theory-circuit-params}; $N=200$).

Lastly, we note a practical consideration: the NLTL circuit contains large superconducting loops formed by paths through the tank circuit and in-line inductors linking ground plane connections.
Large loops increase sensitivity to external magnetic field noise.
In Appendix~\ref{app:path_integral}, we discuss how additional capacitors can mitigate this flux sensitivity by interrupting the large flux loops and provide conditions under which the modified circuit approximates the desired field dynamics.

\begin{table}[t!]
    \centering
    \renewcommand{\arraystretch}{1.3}
    \begin{tabular}{|c|c|c|c|c|c|c|c|}
    \hline
    Field theory & N & $E_C$ & $E_J$ & $E_{JA}$ & $n$ & $b$ & $E_L$  \\
    \hline
    single-well & 200 & 1.94 & 8.2 & $12E_J$ & 10 & $\pi$ & 8.18 \\
    double-well & 200 & 1 & 6 & $2E_J$ & 20 & $\pi$ & 0.18 \\
    \hline
    \end{tabular}
    \caption{\textbf{Proposed circuit parameters for interacting transmission lines.} Top row: parameters for realizing strongly-interacting $\phi^4$ theory with positive coefficients; selected in Section~\ref{sec:scattering}. Bottom row: fluxonium-like parameters for realizing kink/anti-kink states in the deep double-well regime; selected in Section~\ref{sec:kink}. Energetic quantities are presented in units of ($h$~GHz).}
    \label{tab:interacting}
\end{table}

\subsection{Origin of the nonlinearity}

Another nonlinear transmission line is the traveling wave parametric amplifier (TWPA)~\cite{obrien2014, Macklin2015}. 
In TWPAs, the nonlinearity originates from Josephson junctions placed in line (at the equivalent position of $L_\text{line}$ in our device).
The nonlinearity in a TWPA therefore arises from a field-gradient term $(\partial_x \phi_x)^4$, which is small for low-energy (long-wavelength) modes. 

In our NLTL, as in designs that simulate the sine-Gordon model~\cite{Roy2021sinegordon}, the nonlinearity instead arises from Josephson junctions connecting each node to ground. 
The resulting nonlinear term $\phi_x^4$ is therefore a function of field itself.
Accordingly, our device has exceptionally strong interactions between even low-energy particles. 

In the single-well regime, the nonlinearity is enhanced by suppressing the quadratic coefficient through competition between the Josephson junction and the inductor.
For the proposed parameters, the relative anharmonicity of the tank circuit is $\alpha_r = (E_{12} - E_{01})/E_{01} =+$~19.8\%, compared to roughly $-5\%$ for a typical transmon qubit. 
In the double-well regime, the nonlinearity arises from the near-degeneracy of the ground state---a qualitatively different and inherently non-perturbative mechanism.

\subsection{Unboundedness}

Several proposals and demonstrations of lattice gauge theory simulation have considered $\mathbb{Z}_2$ models, which admit a natural qubit encoding~\cite{Zohar2017, Schweizer2019, Charles2024, Mildenberger2025}.
Continuous fields can be represented by digitizing the field amplitude, and the resulting local Hilbert space is then encoded in a multi-qubit register at each site.
Such encodings involve a tradeoff between truncation accuracy, qubit requirements, and circuit depth~\cite{jordan2012quantum}.
Analog proposals for emulating sine-Gordon field theory~\cite{Roy2021sinegordon, Roy2023sinegordon} and pure-gauge compact QED~\cite{Pardo2025} circumvent truncation by encoding the field variable in a superconducting device, leveraging that the phase across a Josephson junction is a compact variable with a sinusoidal potential.
Because this variable is $2\pi$-periodic, such circuits comprising Josephson junctions and capacitors naturally emulate theories with compact field variables.

In contrast, we have presented a transmission line that encodes a continuous and \emph{unbounded} field theory.
As in Ref.~\cite{Roy2021sinegordon}, the continuity of the field variable arises from the analog encoding of the field theory.
It is unbounded because we have placed a linear inductor in parallel with the Josephson junction.
When using $n$-junction superinductors, the field variable becomes $2\pi n$-periodic.
We justify that the field remains effectively unbounded, using the parameters suggested for the positive $\phi^4$ field theory, as follows:
we calculate the phase variance of the first few eigenstates, finding $\langle \varphi^2 \rangle=0.58\pi$, $1.54\pi$, and $2.32\pi$ for the ground, first excited, and second excited states, respectively.
Because $\langle \varphi^2 \rangle\ll 2\pi n$, the phase variable is effectively unbounded.
Meanwhile, the linear inductor additionally serves to suppress charge sensitivity~\cite{manucharyan2012evidence, dipaolo2020tensor, randeria2024dephasing}.

\section{Quantitative comparison of NLTL to $\phi^4$ theory}
\label{sec:quant-non-lin}
In this section we quantitatively compare $\phi^4$ theory and the quantum field theory encoded in the NLTL. 
We confirm that the spectrum of the NLTL faithfully reproduces that of lattice $\phi^4$ theory, that the encoded theory is non-perturbative, and that, with $N=200$ lattice sites, the NLTL is near the continuum limit.
Computations use the discrete Lagrange densities Eq.~(\ref{eq:lattice-phi4-theory}) and Eq.~(\ref{eq:lagrange_NLTL}) for $\phi^4$ theory and the NLTL, respectively.
Throughout this section we consider only the single-well regime, using the parameters given in the top row of Table~\ref{tab:interacting}, since the double-well regime is inherently non-perturbative.

\begin{figure}[b!]
  \centering
  \includegraphics[width=0.38\textwidth]{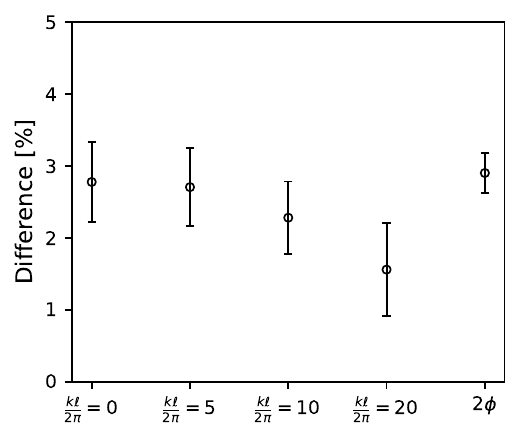}
  \caption{\textbf{Comparing NLTL to $\phi^4$ theory.} Fractional difference between energy levels of the NLTL and $\phi^4$ theory ($\text{Difference} = (E_{\phi^4}-E_{\text{NLTL}})/ E_{\phi^4}$, where $E_{\phi^4},E_{\text{NLTL}}$ respectively denote energies computed in $\phi^4$ and  NLTL). From left to right energies correspond to the $k\ell/2\pi = 0,5,10,20$ single particle energy states then the zero-momentum two-particle state.  }
  \label{fig:comparing-spectra}
\end{figure}

We compute observables in the two systems using a recent adaptation of lattice field theory for finding the static properties of superconducting circuits~\cite{Lin:2025edb}. The approach relies on a path-integral formulation of quantum circuits, which is then solved numerically with Monte Carlo methods. Further details on both the lattice method and the specific calculations presented here are in Appendix~\ref{app:lattice-calc-details}.

We first validate that the NLTL reproduces $\phi^4$ theory.
The NLTL Lagrangian contains extraneous terms of order $\mathcal{O}(\varphi^6)$ and higher from the cosines in Eq.~(\ref{eq:tank_potential}).
We compute the energies of five states: the $k = 2\pi/\ell \times \{0,5,10,20\}$ single-particle states, and the zero-momentum two-particle state.   
As seen in Fig.~\ref{fig:comparing-spectra}, all energy levels agree to within a few percent, indicating close agreement between the two theories. 

Importantly the single-particle rest mass (the energy of the $k=0$ mode) of the NLTL is $7.35(3)\, h\, \text{GHz}$, about $\sim50\%$ higher than that of the free-field transmission line (where it is $5.03\, h\, \text{GHz}$). This shift is produced by the $\phi^4$ interactions and highlights the large nonlinearity.

We next demonstrate that the particle interactions are of non-perturbative strength by comparing the particle mass from lattice field theory to that from perturbation theory.
Perturbative values are computed from lattice (rather than continuum) $\phi^4$ theory at up to third order in ${\lambda}$.
The leading-order perturbative correction to the particle mass squared arises from the 1-loop Feynman diagram
\begin{equation}
\label{eq:one-loop-diagram}
\raisebox{-0.3ex}{%
\begin{tikzpicture}[scale=0.5, line width=1.3pt, baseline={(0,0.4)}]
  \draw (-2,0) -- (2,0);
  \draw (0,0.85) circle (0.85);
  \filldraw (0,0) circle (0.09);
\end{tikzpicture}} = \delta \omega_0^2 = \frac{\tilde{\lambda}\omega_0^2}{4\pi} \frac{K(\frac{4}{(4+r^2)})}{\sqrt{1+r^2/4}}~,
\end{equation}
where $r = \Delta x \omega_0/c = (L_{\text{line}}/L_{\text{tank}})^{1/2}$ is a dimensionless lattice spacing, $\tilde{\lambda} = \hbar \lambda/(\xi c \omega_0^2 )$ is a dimensionless interaction strength, and $K(x) = \int_0^{\pi/2}ds (1 - x \, \text{sin}^2s)^{-1/2}$ is the complete elliptic integral of the first kind. Note that the leading-order mass shift is proportional to Planck's constant. It is therefore a quantum mechanical effect and not simply the result of classical nonlinearities in the transmission line. The higher-order corrections involve a lengthy computation that we outline in Appendix \ref{app:lattice-calc-details}.

\begin{figure}[t!]
  \centering
  \includegraphics[width=0.38\textwidth]{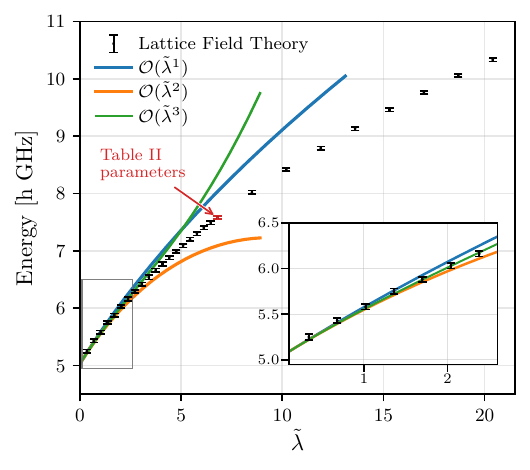}
  \caption{\textbf{Non-perturbative regime.} Comparison between particle masses computed with lattice field theory (black points) and perturbation theory (PT) at first through third order in $\tilde{\lambda}$ (blue, orange,  green lines, respectively), where $\omega_0 = 5.03 \mathrm{GHz}$. 
  The interaction strength produced by Table \ref{tab:interacting} parameters is indicated in red.
  Inset panel shows comparison between perturbative and nonperturbative results near $\tilde{\lambda} = 0$. }
  \label{fig:finding-non-pert}
\end{figure}

In Fig.~\ref{fig:finding-non-pert}, we compare the perturbative and lattice field theory values for the particle mass. As the inset shows, quantitative agreement is found between the two approaches at small values of $\tilde{\lambda}$. 
However, as the interaction strength grows, the values begin to diverge.
At $\tilde{\lambda} \simeq 6.8$, the value produced by Table~\ref{tab:interacting} parameters, the perturbative particle mass shows large order-by-order jumps rather than convergence, indicating a breakdown of perturbation theory. We conclude that the theory is non-perturbative---in fact, the perturbative series should be asymptotic and have a radius of convergence of zero~\cite{Dyson:1952tj}.

\begin{figure}[t!]
  \centering
  \includegraphics[width=0.38\textwidth]{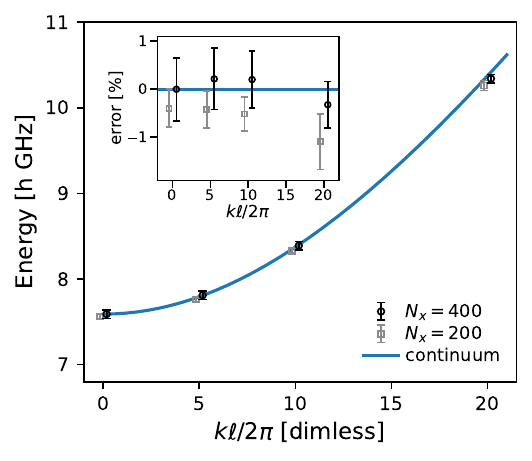}
  \caption{\textbf{Continuum limit.} Computation of the single particle dispersion relation $E(k)$ at Table \ref{tab:interacting} parameters (gray points), at parameters corresponding to 1/2 the lattice spacing (black points) and the continuum limit (blue). Inset shows the fractional error in lattice-computed energies relative to the continuum limit. Points are shifted horizontally for visibility. }
  \label{fig:continuum-limit}
\end{figure}

Finally, we confirm that the NLTL produces a lattice $\phi^4$ theory that is close to the continuum limit. 
We compute single-particle energy levels for several values of the momentum ($k = 2\pi/\ell \times \{0,5,10,20\}$). 
Then, we double the number of lattice sites to $N=400$, halve the lattice spacing, and recompute the same energy levels. As shown in the inset of Fig. \ref{fig:continuum-limit}, the energy levels change by around $1\%$ or less, confirming close agreement with the continuum limit.
Furthermore, one can show that the \textit{expected} continuum limit is approached. While this is an interacting quantum field theory and thus the energy levels of the system are not obvious \textit{a priori}, Lorentz invariance of the continuum theory implies that the single-particle dispersion relation is $E(k)= \sqrt{E_0^2+(\hbar k c)^2}$, where $E(k)$ is the total energy, $E_0$ is the rest mass energy, $k$ is the wavenumber, and $c$ is the effective speed of light. In Fig.~\ref{fig:continuum-limit}, we plot this curve with a rest mass energy of $E_0 = 7.59~h$~GHz (the value extracted from the $N_x = 400$ lattice calculations) and $c=\Delta x/\sqrt{C L_{\text{line}}}$. The close agreement between this prediction and the numerical calculations confirm that the chosen parameters achieve a lattice field theory near the correct continuum limit.

\section{Inelastic Wave Packet Scattering}\label{sec:scattering}

\begin{figure}[t!]
  \centering
  \includegraphics{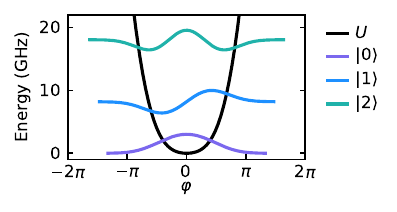}
  \caption{\textbf{Energy spectrum of one tank circuit in the single-well regime}. The potential is indicated by the black line, and the wavefunctions of the ground state and first two excited states are shown offset by their energy. Tank circuit parameters are given in the top row of Table~\ref{tab:interacting}.}
  \label{fig:phi4_tank_spectrum}
\end{figure}

Particle production is a distinctive feature of relativistic theories. 
We describe here a scattering experiment between two particle wave packets within the $\phi^4$ field theory, and we describe how to detect the outgoing particles. 
Due to the strong interactions, we expect the incoming wave packets to scatter into several lower-energy outgoing wave packets. 
This feature is called particle production in high-energy physics and down-conversion in optics. 
Such a two-to-many-particle scattering experiment is possible in our setup.
The parameters of the transmission line should here be chosen to produce a single-well potential (positive $\varphi^2$ and $\varphi^4$ coefficients; top row of Table~\ref{tab:interacting}).
Operating near the quarton regime provides significant anharmonicity~\cite{Ye2021QuartonPRL}.
The energy spectrum of a single tank circuit is shown in Fig.~\ref{fig:phi4_tank_spectrum}.

\subsection{State Preparation}\label{sec:prep}

To study particle scattering within the interacting field theory, we must prepare suitable initial states in the transmission line. 
We would like to prepare two (or more) particle states comprising wave packets at opposing ends of the transmission line so that the wave packets propagate to the center of the transmission line and collide.
Superconducting transmission lines naturally lend themselves to spectroscopic measurement, but these experiments create states with {\em nondeterministic} particle number~\cite{Fitzpatrick2017dissipative, Kim2021ssh}. 
On the other hand, intentional state preparation has become routine in quantum simulation work aboard superconducting quantum computers~\cite{Mi2022timecrystal, Rosen2024}, however creating wave packets requires lengthy digital algorithms~\cite{Farrell2025wstates}. 

In this section, we propose coupling several qubits to the transmission line, enabling deterministic preparation, via an analog protocol, of wave packet states for scattering experiments.
Consider first the free-field transmission line.
We capacitively couple a flux-tunable transmon qubit to each end of the transmission line. 
The qubit energies are selected to be below the passband of the transmission line (e.g., for $5~h~\mathrm{GHz}$ particle mass, at $4.5~h~\mathrm{GHz}$) to prevent unwanted interaction. 
To begin each experimental sequence, a single photon is loaded into each qubit via a microwave $\pi$-pulse. 
Then, both photons are simultaneously transferred into the transmission line to form a converging pair of propagating one-photon wave packets.
This transfer is accomplished by delivering a parametric flux pulse to the qubits~\cite{kannan2023directional, joshi2023resonance, almanakly2025entanglement}.
The frequencies of the parametric pulses are the difference between the energies of the desired wave packets and the qubits, and the spatial profiles of the wave packets are determined by the amplitude envelope of the pulses.
Initial states with more particles may be realized via the same protocol using more qubits.

\begin{figure}
    \centering
    \includegraphics{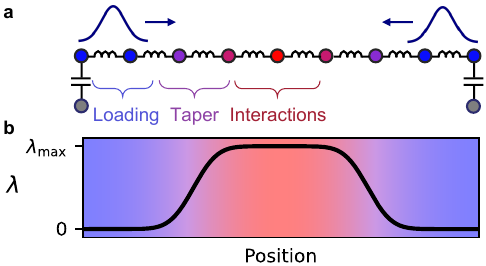}
    \caption{\textbf{Representation of wave packet preparation.} \textbf{(a)} Wave packets are loaded into the loading zones (tank circuits with $\lambda=0$; blue dots) from qubits (grey dots) that are capacitively coupled to the terminal tank circuits. Interactions are adiabatically introduced as the wave packets travel through the taper segments (purple dots), and the wave packets collide in the interaction zone ($\lambda \neq 0$; red dots). For clarity, an abbreviated number of tank circuits are shown. \textbf{(b)} A schematic of the interaction strength $\lambda$ along the transmission line.}
    \label{fig:stateprep}
\end{figure}

Now we wish to prepare wave packets of the interacting field theory. 
As we have no a priori solution of $\phi^4$ in the non-perturbative regime, we do not know what those wave packets look like. 
Instead, let us prepare wave packets of the non-interacting field theory, and then adiabatically introduce the interaction prior to their collision.
We take inspiration from the tapered impedance transformers that allow microwaves in a waveguide to pass between regions of differing impedance with little reflection~\cite{klopfenstein1956taper, pozar2011microwave, Kim2021ssh, Zhang2023metamaterial}.
We divide our transmission line into several sections, as shown in Fig.~\ref{fig:stateprep}.
At both ends of the transmission line, the nonlinearity is set to $\lambda=0$. 
The center of the transmission line has strong interactions; this is where the two wave packets will scatter. 
These regions are connected by tapers, where the interaction smoothly varies site-to-site.
The interaction is varied by adjusting $E_J$ and $E_{JA}$ to exchange inductive energy stored in the Josephson junction to the linear inductor, so that the $\phi^4$ coefficient in Eq.(\ref{eq:tank_potential}) is reduced while holding the quadratic terms in the constant.
The length of the particle loading, taper, and interaction segments should all exceed the wave packet size.
In this way, we may deterministically prepare propagating wave packets of the linear field theory in each end of the transmission line and adiabatically introduce interactions as they propagate, culminating in wave packet collision within the interacting field theory.

\subsection{Measurement}

After state preparation, each experimental sequence consists of a free evolution period followed by a measurement. 
At the conclusion of the free evolution period, particle propagation may be frozen by rapidly ramping the magnetic flux bias of every second lattice site from $b=\pi$ to $b=0$. 
Doing so increases the excitation energy of every second lattice site by \SI{7.5}{GHz}, far exceeding the inter-site inductive coupling energy.
After the flux ramp, the coupling between tank circuits is therefore strongly suppressed, halting further population transfer.
The population of individual sites may then be determined by standard single-shot superconducting readout---by interrogating a dispersively coupled readout resonator~\cite{Blais2004, Wallraff2004}.
We note that while particle number may not be conserved within the interaction zone during the free evolution period, it is well-defined during readout when adjacent tank circuits are effectively decoupled.

The full experimental sequence---state preparation, free evolution, and readout---is then repeated after a waiting period that allows dissipation of photons from the transmission line.
Particle number observables are reconstructed from the statistics of many measurement shots.
Repeating the process for a range of free evolution times yields information about the real-time dynamics of the post-collision wave packet.
For example, particle production and velocities may be inferred from the wavefront~\cite{Farrell2025wstates, hartnett2026fermihubbard}. 
Interrogating resonators coupled to several tank circuits provides simultaneous readout at multiple spatial locations along the transmission line. 
Two-point correlation functions may be determined from the joint measurement outcomes.

An alternative measurement scheme would employ a quantum flux parametron coupled to an in-line inductor by a mutual inductance~\cite{Grover2020}.
This setup would directly measure $\varphi_{i+1}-\varphi_{i}$, corresponding to the discrete spatial derivative of $\phi$. 

\section{Kink--Antikink Scattering}\label{sec:kink}

\begin{figure*}
    \label{fig:kink_prep}
    \subfloat{\label{fig:8a}}
    \subfloat{\label{fig:8b}}
    \subfloat{\label{fig:8c}}
    \subfloat{\label{fig:8d}}
    \centering
    \includegraphics{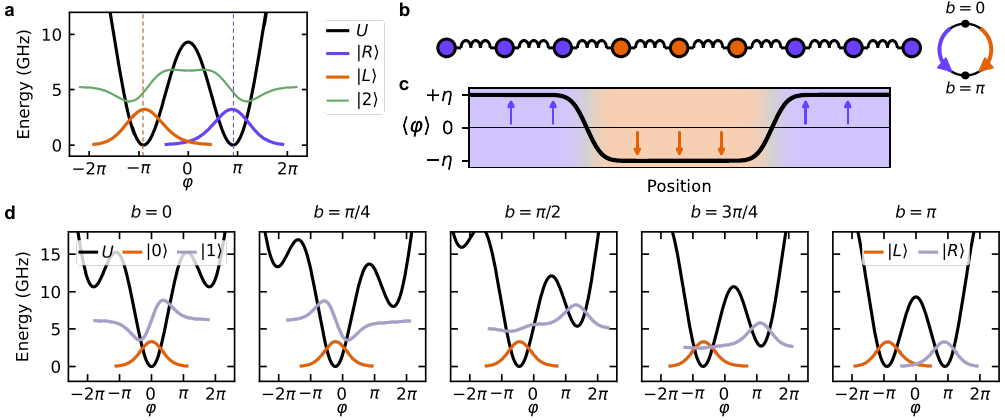}
    \caption{\textbf{Kink-antikink state preparation.} \textbf{(a)} The potential energy (black) of one tank circuit in the double-well regime. The right (orange) and left (violet) well states are the symmetric and anti-symmetric combinations of the true ground and first excited states, which are nearly degenerate. The next excited state (green) is much higher in energy. The values of $\pm \eta$ are indicated by the vertical dashed lines. \textbf{(b)} To prepare a kink-antikink state, a central segment of the transmission line is ramped from an external flux of $b=0$ to $\pi$ in the positive direction, and the remainder in the negative direction. \textbf{(c)} Initially, $\langle \phi \rangle=0$, and after the ramp, $\langle \phi \rangle=-\eta$ in the center segment and $+\eta$ elsewhere. The boundaries (grey shading) are the kink and the antikink. \textbf{(d)} The instantaneous eigenstates of one tank circuit during a ramp from $b=0$ to $\pi$ to establish the left well state. The wavefunctions of the ground (orange) and first excited (light purple) states are shown, offset by their energies.}
\end{figure*}

The negative-mass regime provides a complementary route to nonperturbative real-time dynamics. Consider parameters for the tank circuit near the fluxonium regime, for example the parameters presented in Table~\ref{tab:interacting}. In this regime, when $b=\pi$, the onsite potential forms a deep double well with a large (compared to the kinetic energy) barrier between the two wells. 
The two lowest eigenstates are nearly degenerate and are respectively symmetric and antisymmetric under $\varphi\to-\varphi$.
Their equal-weight superpositions define states localized predominantly in the left and right wells, denoted by $|L\rangle$ and $|R\rangle$ (Fig.~\ref{fig:8a}).

These two states correspond to the two approximately degenerate vacua of the  low-energy field theory, which we denote by $\phi=\pm \eta$. A kink $K$ is a domain wall interpolating from one vacuum to the other. Conversely, an antikink $\bar{K}$ interpolates in the opposite direction. Thus, a neutral kink--antikink pair in a finite transmission line is prepared by initializing a spatial pattern of vacua,
\begin{equation}
    +\eta \quad \longrightarrow \quad -\eta \quad \longrightarrow \quad +\eta ,
\end{equation}
where the first interface is the kink and the second interface is the antikink (Fig.~\ref{fig:8c}). 
At low energy, the kink and antikink are topological distortions of the field with localized energy density.

We aim to initialize a kink--antikink pair in the transmission line and to allow them to scatter. 
This experiment amounts to initializing a corresponding spatial pattern of $|L_i\rangle$ and $|R_i\rangle$ (Fig.~\ref{fig:8b}) followed by a free evolution period.
As the dynamics within transmission line is, to the extent possible, unitary, we expect the kink--antikink annihilation process to result in energetic emission.
Our objective is to observe this emission. 
As in the target theory, higher energy levels beyond the vacua are available and may become occupied following kink--antikink scattering.

\subsection{Preparing Kink--Antikink States}\label{sec:kink_prep}

Local control of the flux bias enables intentional and spatially patterned preparation of the two vacua. Consider initializing the transmission line with $b=0$ for all nodes. Here, the tank circuit potential is effectively a single well centered at $\varphi=0$, and the transmission line will equilibrate to its ground state, namely every node in the nondegenerate ground state $|0\rangle$ of the single well. We now wish to quickly (i.e. faster than the thermodynamic equilibration time of the transmission line $T_1$) ramp to the symmetric double-well point $b=\pi$ in a manner that prepares the desired vacuum profile. Consider, momentarily, an isolated tank circuit. If $b$ is adiabatically ramped from zero towards $+\pi$, the center of the single well moves from $\varphi=0$ towards $\varphi=-\eta$ and the state $|0\rangle$ will follow (Fig.~\ref{fig:8d}). Near $b=+\pi$, as the double well is established, the ramp will naturally become diabatic because the coupling between the two vacua vanishes. Therefore, the ramp establishes the left double well vacuum state. Conversely, ramping $b$ towards $-\pi$ moves the single well from $\varphi=0$ towards $\varphi=+\eta$ and establishes the right double well vacuum state.

Reconsidering the full transmission line, we therefore can initialize a kink--antikink state by ramping $b$ in a central segment of the transmission line to $-\pi$, establishing $+\eta$, and to $+\pi$ in the remainder, establishing $-\eta$. At the conclusion of the ramp, the transmission line has a uniform symmetric double-well potential, and a kink and antikink has been formed at the left and right boundaries of the central segment, respectively.
The field profiles of the kink and antikink are discussed in Appendix~\ref{app:profile}.
The subsequent dynamics allow the kink and antikink to move, spread, and interact, potentially leading to collision and annihilation.

\subsection{Expected Dynamics}
The inductive coupling penalizes differences between neighboring field values, giving an energy cost to domain walls.
The local charging term supplies the quantum dynamics, including tunneling between the two wells, which allows the domain walls to move.
Their evolution can be monitored through the topological-charge profile,
\begin{equation}
    q_i(t)=\frac{\varphi_{i+1}(t)-\varphi_i(t)}{2\eta}.
\end{equation}
Initially, $\langle q_i(t)\rangle$ exhibits two localized peaks of opposite sign at the kink and antikink positions.
During evolution, these peaks may move and broaden, reflecting changes in the domain-wall profiles and quantum spreading of their positions.
The evolving kink and antikink may meet and scatter.

The scattering event is then observed
as the motion, collision, and possible disappearance, reflection, or
re-emergence of these topological-density peaks. If the kink and antikink
annihilate, the energy initially stored in the two domain walls may be converted into propagating non-topological excitations of the scalar field, conventionally called mesons in $\phi^4$ theory.
Their production may be inferred from population in $|2\rangle$ and higher levels after ramping the external flux bias back to the single-well configuration ($b=0$), using the readout protocol described below.
In particular, kink--antikink annihilation into mesons would give a direct real-time probe of nonperturbative dynamics in the
broken-symmetry $\phi^4$ theory. 

Whether the collision can produce these scalar excitations depends on the available energy and the excitation spectrum of the transmission line.
A relevant local energy scale is the gap between the two lowest-energy states of an isolated tank circuit and the next level, $|2\rangle$, approximately $\SI{5.2}{GHz}$ (green line in Fig.~\ref{fig:8a}).
State preparation creates an approximately step-like kink with energy $E_K=(E_L/2)(2\eta)^2$.
For the parameters in Table~\ref{tab:interacting}, the initial configuration therefore stores an estimated energy $2E_K=5.9\, h\, \mathrm{GHz}$.
The available energy thus exceeds the isolated-circuit excitation scale, suggesting that this parameter regime is promising for exploring meson production.
Varying $E_L$ changes the energy stored in the initial domain walls, providing a way to explore how the collision dynamics depend on the available energy.

\subsection{Measuring the Dynamics}

We wish to study the real-time scattering of $K$ and $\bar K$ and determine the emission products, e.g. mesons, a long-lived oscillating bound state, or outgoing solitonic excitations. 
This process is intrinsically nonperturbative: the incoming kink and antikink are extended field configurations connecting distinct vacua and are not finite-order perturbative excitations around either vacuum.

The local value of the field is determined by measuring $\langle \varphi_i(t)\rangle$. To do so, we reverse the external flux ramp at time $t$ after state preparation.
That is, we ramp $b$ from $-\pi$ to zero in the central segment and from $+\pi$ to zero elsewhere. 
Once returned to $b=0$ (again provided an adiabatic ramp), sites where the initial vacuum state were maintained will return to $|0\rangle$.
Sites that have changed between the two vacua $\pm \eta$ during the free evolution period (that is, sites that have switched from $|L_i\rangle$ to $|R_i\rangle$ or vice-versa) will instead be brought to the first excited state at $b=0$, $|1\rangle$.
Particle production will result in $|2\rangle$ (or higher). 

The final state may be detected using standard superconducting qubit readout. At $b=0$, $|1\rangle$ is a plasmon excitation of $|0\rangle$ (like the logical $|1\rangle$ state of a transmon qubit). Readout may therefore be accomplished by interrogating a capacitively interacting, dispersively coupled readout resonator.

Repeating this sequence for many values of $t$ reconstructs the real-time collision dynamics from snapshots of single-point $\langle \varphi_i(t)\rangle$ and equal-time two-point $\langle \varphi_i(t)\varphi_j(t)\rangle$ functions. 
From these local measurements one can reconstruct the topological-charge profile. 
Successful kink--antikink preparation may be verified by two separated peaks in $q_i$ at $t=0$ with opposite signs, and their subsequent evolution tracks the collision.
The energy associated with these configurations can be characterized by the local energy per site,
\begin{equation}
  \epsilon_i(t)=\bigg\langle
    \frac{C\varphi_0^2}{2}\dot{\varphi}_i^2
    +U(\varphi_i,b)
    +\frac{E_L}{2}
      (\varphi_{i+1}-\varphi_i)^2
  \bigg\rangle_t
\end{equation}
where $\varphi_0=\hbar/(2e)$ is the reduced flux quantum and each coupling bond is assigned to its left site.
Relative to the uniform vacuum, the initial configuration stores excess energy near the two domain walls.

\section{Conclusion}

We have presented a superconducting microwave transmission line that is an analog quantum emulator of strongly interacting $\phi^4$ field theory.
Like particle colliders, the transmission line provides information about the dynamics of particle scattering.
The real-time dynamics of strongly interacting scalar field theories, including the $\phi^4$ model, are not analytically tractable, and their simulation presents formidable challenges for classical computational methods and for digital quantum computers.

To non-perturbatively capture the physics of high-energy scattering on classical computers, the continuous field amplitude must be digitized into a finite number of discrete levels at each spatial lattice site. 
High-energy collisions and the resulting multiparticle production can generate large fluctuations in the field. 
Accurately resolving these highly excited states requires a wide field cutoff and a fine discretization grid, so an overly restrictive local truncation can introduce artifacts. In classical tensor network algorithms, such as those based on Matrix Product States (MPS), the computational cost scales with this local dimension. Entanglement generated during scattering can further increase the required bond dimension, making sufficiently complex real-time dynamics computationally demanding.

Digital quantum algorithms avoid this bottleneck by encoding scalar fields in qubit registers whose size grows only logarithmically with the local dimension~\cite{jordan2012quantum}, but realizing the field dynamics requires quantum gate sequences whose cost depends on the encoding, target accuracy, and hardware connectivity.
On devices with restricted connectivity, additional routing operations further increase circuit depth and error accumulation.
The transmission line proposed here takes a complementary approach: each lattice field variable is represented directly by a continuous, unbounded circuit coordinate, and the interactions and time evolution are generated by the circuit Hamiltonian.
This avoids digitizing the field amplitude.
The accuracy is instead governed by circuit coherence, parameter control, and how closely the implemented Hamiltonian approximates the target field theory.
This architecture therefore offers a promising route to scattering experiments without the circuit-depth requirements of a digital implementation.

\begin{acknowledgments}
N.C.W.'s contributions were supported by the U.S. Department of Energy, Office of Science under grant Contract Numbers DE-SC0011090 and DE-SC0021006, the Simons Foundation grant 994314 (Simons Collaboration on Confinement and QCD Strings), and the U.S. Department of Energy, Office of Science, National Quantum Information Science Research Centers, Co-Design Center for Quantum Advantage under Contract No. DE-SC0012704.

J.L.'s contributions were supported by
the U.S. Department of Energy, Office of Science, Office of Nuclear Physics through Contract No. DE-AC02-06CH11357, and by the U.S. Department of Energy, Office of Science, Office of Nuclear Physics, Early Career
Award through Contract No. DE-SCL0000017.

J.B. is supported by a Feodor Lynen Research Fellowship from the Alexander von Humboldt Foundation.

This material is based on research sponsored by the Air Force Office of Scientific Research under agreement number FA9550261B128. The U.S. Government is authorized to reproduce and distribute reprints for Governmental purposes notwithstanding any copyright notation thereon.

Claude was used for the perturbative calculation, debugging of code, the generation of some figures and some editing. The authors take full responsibility for the correctness of the reported results.
\end{acknowledgments}

\appendix


\section{Numerical calculations}\label{app:lattice-calc-details}
In this section we detail our lattice field theory and perturbative calculations. For further information on the lattice method see \cite{Lin:2025edb}. Lattice field theory is a method for computing the static properties (energy levels and matrix elements) of arbitrary quantum mechanical many-body systems. These are extracted from thermally-averaged Euclidean-time correlation functions $C(t) = \langle \mathcal{O}(t) \mathcal{O}(0)\rangle= \text{tr}(e^{-\beta H}\mathcal{O}(t) \mathcal{O}(0))/\text{tr}(e^{-\beta H})$ computed in a path integral formalism. Here $\mathcal{O}(t) = e^{-H t / \hbar} \mathcal{O} e^{H t / \hbar}$ is a Euclidean time evolved operator, $H$ is the Hamiltonian of the system, and $\beta$ is a fictitious inverse temperature. Static quantites are extracted from multi-exponential fits to correlation functions, which can be re-written as
\begin{align}
\label{eq:twopointspec}
    C(t) = Z^{-1}\color{black}\sum_{nm}e^{-\beta E_n} & e^{-(E_m-E_n)t/ \hbar }  |\langle n | \mathcal{O} | m \rangle|^2  
\end{align}
where $E_n, |n\rangle$ are respectively many-body energy levels and eigenstates and $Z= \text{tr}(e^{-\beta H})$. Path-integral expressions for correlators are obtained by trotterizing the Euclidean-time evolution of operators into $N_t$ steps of length $\Delta t = \hbar \beta/N_t$. Up to $O(\Delta t^2)$ Trotterization errors, one finds 
\begin{align}
    C(t) = \frac{\int D \phi ~ e^{-S(\phi)/\hbar }   \mathcal{O}(t) \mathcal{O}(0)}{\int D \phi ~ e^{-S(\phi)/\hbar }} ~,
\end{align}
where $S$ is the Euclidean action of the circuit, given by
\begin{equation}
    S =  \frac{1}{2} \varphi_0^2 \sum_{t, x, x'} \Delta t  ~\dot{\phi}_{x,t}C_{xx'}\dot{\phi}_{x,t} + \sum_{t,x} \Delta t ~U(\phi_{x,t}).
\end{equation}
Here $\phi_{xt}$ is the node flux on site $x$ and timeslice $t$, which is collected into a ``field configuration" $\phi = \{\phi_{xt}\}$, $\dot{\phi}_{x,t} = (\phi_{x,t+1}-\phi_{x,t})/\Delta t$, $\varphi_0=\hbar/(2e)$ is the reduced flux quantum, $C_{xx'},U$ are respectively the capacitance matrix and potential of the circuit, and $D\phi = \prod_{t=1}^{N_t}\prod_{x=1}^N d\phi_{xt}$ product of node-flux variables on all nodes and timeslices. To numerically compute the required path integrals, Markov Chain Monte Carlo methods are used. Many field configurations distributed according to the probability distribution $p(\phi) = e^{-\frac{S(\phi)}{\hbar} }/ \int D\phi ~e^{-\frac{S(\phi)}{\hbar}}$ are collected, and correlation functions are stochastically estimated with direct averaging. The finite statistics of the Markov chain results in a statistical error on all reported observables, which vanishes at a rate dictated by the central limit theorem.

\begin{figure}[t!]
  \centering
  \includegraphics[width=0.38\textwidth]{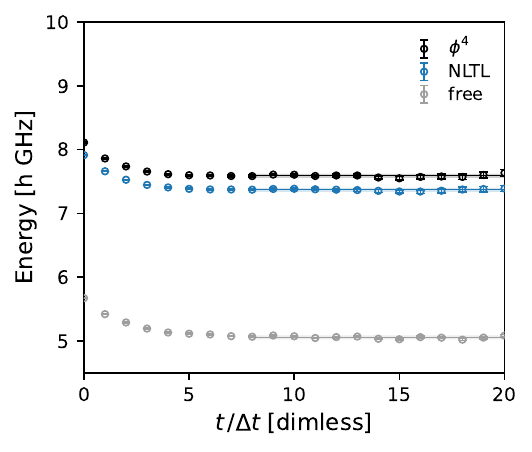}
  \caption{\textbf{Effective mass plot example.} Calculation of the particle rest mass in $\phi^4$ theory (black), the non-linear transmission line (blue), and the free-field transmission line (gray). Dashed lines indicate the value of the particle rest masses. The horizontal axis is the number of timeslices advanced in the correlation function. }
  \label{fig:eff_mass_example}
\end{figure}

Results for five energy levels are reported in Section \ref{sec:quant-non-lin}. These are extracted from correlation functions of $\mathcal{O} = \sum_x e^{-i k x } \phi_x$ for $k=2\pi/\ell \times\{0,5,10,20\}$, which create single particle momentum eigenstates, and $\mathcal{O} = (\sum_x \phi_x)^2$, which creates a two-particle state. For long times,  correlation functions plateau to the leading exponential in Eq.~(\ref{eq:twopointspec}), and the \textit{effective mass}
\begin{equation}
\label{eq:discrete-eff-freq}
    E(t) = -\frac{\hbar}{\Delta t} \text{log}\big[ C(t+\Delta t)/C(t) \big]~,
\end{equation}
plateaus to a constant. Effective mass plots are useful diagnostics for extracting energy levels. Examples for $\mathcal{O} = \sum_x\phi_x$ are shown in Fig.~\ref{fig:eff_mass_example}. Typical behavior is seen, with ``excited state contamination" at small times with an eventual plateau.

The results from several lattice calculations are presented in Section \ref{sec:quant-non-lin}. In all cases, observables are computed from Markov chains of $128,000$ field configurations generated using the hybrid Monte Carlo algorithm  \cite{DUANE1987216}, and single exponential fits to correlators were found to be sufficient to extract  energy levels. Statistical errors were estimated with bootstrap resampling. For all presented lattice calculations, both finite volume (i.e. $\ell \neq \infty$) and $\Delta t$ errors were examined and found to be negligible compared to the statistical errors. 

Fig. \ref{fig:comparing-spectra} is generated by comparing an NLTL with $N=200$ unit cells with a lattice $\phi^4$ theory with the same number of lattice sites. For both theories, parameters are taken to be equal to the top row of Table~\ref{tab:interacting} and using the mappings Eq.~(\ref{eq:int_theory_mapping}). In calculations $N_t=400$ and $\Delta t = 2.5 \times 10^{-3}$ picoseconds.

Fig.~\ref{fig:finding-non-pert} considers lattice $\phi^4$ theory, with bare values of $\omega_0$ and $v$ held fixed to those produced by Table~\ref{tab:interacting} and $\lambda$ swept across. This choice represents a fixed free theory with variable interaction strength. In these calculations $N=200$, $N_t=400$, and $\Delta t = 2.5 \times 10^{-3}$ picoseconds.

Fig.~\ref{fig:continuum-limit} considers $\phi^4$ theory at two different values of the lattice spacing. A reference set of $\omega_0, v,\lambda$ on an $N=200$ system are produced by taking parameters equal to the top row of Table~\ref{tab:interacting} and using the mappings in Eq.~(\ref{eq:int_theory_mapping}). Then, the number of lattice sites is doubled to $N=400$, while $v,\lambda$ are held fixed. In order that the physical mass of the particle is held fixed, the bare value $\omega_0$ is reduced to roughly $0.8$ its value on the $N=200$ system. In the $N=200$ calculation, $N_t=400$ and $\Delta t = 2.5 \times 10^{-3}$ picoseconds; in the $N=400$ calculation $N_t$ was doubled and $\Delta t$ was halved.

The perturbative calculations were performed as follows. On a finite lattice, the energy of the zero-momentum single particle state $|k_0\rangle = a^{\dagger}_{k=0}|0\rangle$ was computed to third order in non-degenerate perturbation theory.  This is accomplished by expressing the energy
\begin{equation}
    E(\lambda) = E_0 + \lambda E_1 + \lambda^2 E_2 + \lambda^3 E_3 + \mathcal{O}(\lambda^4)~,
\end{equation}
where $E_0 = \hbar \omega_0$, then numerically computing 
\begin{align}
    E_1 & = \langle k_0| V |k_0\rangle \nonumber \\
    E_2 & = \sum_{k\neq k_0}\frac{|\langle k_0 | V |k\rangle |^2}{\epsilon(k_0)-\epsilon(k)} \nonumber \\
    E_3 & = \sum_{k,k'\neq k_0}  \frac{\langle k_0 | V |k\rangle \langle k | V |k'\rangle  \langle k' | V |k_0\rangle}{(\epsilon(k_0)-\epsilon(k))(\epsilon(k_0)-\epsilon(k'))} \nonumber \\
    & \qquad \qquad - \langle k_0| V |k_0\rangle\sum_{k\neq k_0}\frac{|\langle k_0 | V |n'\rangle |^2}{(\epsilon(k_0)-\epsilon(k))^2}~.
\end{align}
Here $\epsilon(k)^2=(\hbar \omega_0)^2+\frac{4 \hbar^2c^2}{\Delta x^2}\text{sin}^2(k\Delta x/2)$ is the lattice free particle dispersion relation, $|k\rangle = a_k^{\dagger}|0\rangle$ are free-particle momentum eigenstates, and $V = \xi \lambda / 4! \sum_x \Delta x :\phi_x^4:$ is a normal-ordered interaction. The curves in Fig.~\ref{fig:finding-non-pert} are obtained by taking the infinite volume limit of these computations, then plotting the energy order-by-order. Note that the infinite volume limit of $E_1$ agrees with the one-loop expression for the self-energy in Eq.~(\ref{eq:one-loop-diagram}).

\section{Field profile of the kink}
\label{app:profile}

For kink and antikink centers at positions $X_K<X_{\bar K}$, the target mean-field profile is
\begin{equation}
    \langle \phi(x,0)\rangle \simeq
    \eta\left[
    \tanh\left(\frac{x-X_K}{\zeta}\right)
    -
    \tanh\left(\frac{x-X_{\bar K}}{\zeta}\right)
    -1
    \right],
    \label{eq:kink_antikink_profile}
\end{equation}
where $\zeta$ is the kink width. 
We do not expect the flux bias ramp (Section~\ref{sec:kink_prep}) to faithfully reproduce this mean-field profile.
Were the tank circuits isolated from one another, the flux bias ramp would produce a discrete jumps between $\phi=-\eta$ (at sites where $b$ is ramped from $0$ to $+\pi$ and $\phi=+\eta$ (where $b$ is ramped to $-\pi$.
This scenario would yield a kink and antikink with $\zeta \rightarrow 0$.

However, due to the inductances $L_\mathrm{line}$ between tank circuits, adjacent tank circuits will interact throughout the ramp.
These interactions will lead to dynamical spreading of the kink and antikink during state preparation, and the resulting kink profile may not match the mean-field profile Eq.~(\ref{eq:kink_antikink_profile}). 
The ramp will nevertheless establish a kink-antikink profile because $K$ and $\bar K$ are topological defects and must form regardless of the exact field profile.
Furthermore, spreading of the kink and antikink during initialization can be minimized by completing the flux bias ramp in a timescale that is short compared to the dynamics, which is straightforward using standard fast-flux control techniques. 

\section{Breaking large flux loops}\label{app:break}

\begin{figure}[t!]
  \centering
  \includegraphics[width=0.4\textwidth]{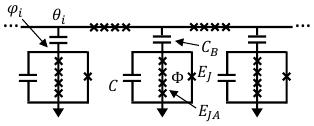}
  \caption{\textbf{Transmission line with flux loop breaks}. Capacitors $C_B$ are added between the tank circuits and the transmission line nodes to interrupt galvanic connection. The capacitors create additional nodes in the circuit defined by the node flux variables $\theta_i$.}
  \label{fig:flux_breaks}
\end{figure}

\begin{figure}
    \centering
    \includegraphics{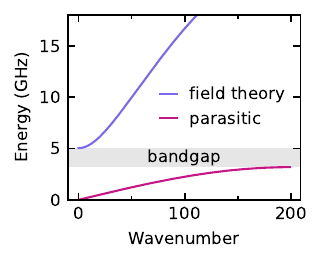}
    \caption{Dispersion of the free-field transmission line with additional capacitors to break unwanted flux loops. Auxiliary qubits for loading particles into the transmission line may be situated within the bandgap.}
    \label{fig:dispersion_full}
\end{figure}

The transmission line described in Fig.~\ref{fig:interacting-field} has numerous coupled flux loops---galvanically uninterrupted loops in the circuit---formed by the path through in-line inductors, the tank circuit inductors, and the connection to ground in each tank circuit. 
Having large flux loops increases the sensitivity of the device to magnetic noise.

To mitigate this flux sensitivity, we introduce additional capacitors to the transmission line that interrupt the large flux loops, as shown in Fig.~\ref{fig:flux_breaks}.
The additional capacitors have capacitance $C_B$ and create a second circuit node per unit cell.
The terms in the circuit Lagrangian that comprise the desired lattice $\phi^4$ field theory are divided between these two nodes, which are coupled by the Lagrangian terms $\sum_i\frac{1}{2}C_B\varphi_0^2 (\dot\varphi_i-\dot\theta_i)^2$ representing the capacitive energy. 
By way of intuition, large $C_B$ penalizes relative motion of $\varphi_i$ and $\theta_i$, so the two nodes move together at the frequencies of the field dynamics.

In Appendix~\ref{app:loops}, we show by change-of-coordinates that the transmission line hosts two coupled modes.
One expresses the desired $\phi^4$ field theory; the second is a gapless parasitic mode.
The single-particle spectrum, determined by exact diagonalization, is shown in Fig.~\ref{fig:dispersion_full} using additional capacitors of energy $E_{CB}/h=\SI{39}{MHz}$, where the capacitive energy is $E_{CB}=e^2/2C_B$.
In Appendix~\ref{app:path_integral}, using a path-integral formulation, we formalize that the effective theory preserves the original $\phi^4$ field theory in the single-well regime by showing that the perturbation on the desired mode by the parasitic mode is suppressed by large $C_B$.

We note that the remaining flux loops in the transmission line are those formed by the Josephson junction and inductor within each tank circuit.
These loops are necessary to adjust the potential energy terms in the field theory by threading intentional external fluxes $b$. 
As these loops are physically small, they are less sensitive to magnetic field noise. 
Experimentally, $b$ may be adjusted by individual on-chip flux bias lines nearby each loop and/or using a global flux bias provided by a nearby coil.

\section{Breaking flux loops: Circuit Lagrangian}\label{app:loops}

The additional capacitors $C_B$ have the effect of splitting the terms in the $\phi^4$ field theory Lagrangian between two node flux degrees of freedom per tank circuit, $\varphi_i$ and $\theta_i$, that are coupled by the additional capacitors. 
This may be seen by the Lagrangian of the broken free-field transmission line $L = \sum_{i=1}^{N} L_i$, where
\begin{equation}
    \begin{split}
    L_i = &\varphi_0^2\Bigg( \frac{C}{2}\dot{\varphi}_i^2 + \frac{C_B}{2}(\dot{\varphi}_i - \dot{\theta}_i)^2 - \frac{1}{2L_{\text{tank}}}  \varphi_i^2 \\  &-\frac{1}{2L_{\text{line}}}  (\theta_{i+1}-\theta_i)^2 \Bigg).
    \end{split}
\end{equation}
We next introduce the coordinates $\chi_i = \theta_i-\varphi_i$, which represent the node fluxes across the additional capacitors.
Parameterized by $\varphi_i$ and $\chi_i$, the local Lagrangian becomes:
\begin{equation}
    \begin{split}
    L_i = &\varphi_0^2\Bigg( \frac{C}{2}\dot{\varphi}_i^2 + \frac{C_B}{2}\dot{\chi}_i^2 - \frac{1}{2L_{\text{tank}}}  \varphi_i^2 \\
    &-\frac{1}{2L_{\text{line}}}  (\varphi_{i+1}+\chi_{i+1}-\varphi_i-\chi_i)^2  \Bigg).
    \end{split}
\end{equation}
Defining the dimensionless conjugate charges $n_i=C\varphi_0^2\dot{\varphi}_i/\hbar$ and $m_i=C_B\varphi_0^2\dot{\chi}_i/\hbar$,
the Hamiltonian is
\begin{equation}\label{eq:h_broken}
\begin{split}
H=\sum_i\Bigg[
    &4E_C n_i^2
    +\frac{\varphi_0^2}{2L_{\text{tank}}}\varphi_i^2
    +\frac{E_L}{2}(\varphi_{i+1}-\varphi_i)^2 \\
    &+4E_{CB}m_i^2
    +\frac{E_L}{2}(\chi_{i+1}-\chi_i)^2 \\
    &+E_L(\varphi_{i+1}-\varphi_i)
          (\chi_{i+1}-\chi_i)
\Bigg]
\end{split}
\end{equation}
with $E_C=e^2/(2C)$, $E_{CB}=e^2/(2C_B)$, and $E_L=\varphi_0^2/L_{\text{line}}$, where $\varphi_0=\hbar/(2e)$ is the reduced flux quantum.
The terms involving only $\varphi_i$ and $n_i$ are exactly the Hamiltonian of the original transmission line. The remaining terms describe the additional degree of freedom $\chi_i$ and its coupling to $\varphi_i$.
The nonlinear tank potential is restored by replacing $\varphi_0^2\varphi_i^2/(2L_{\text{tank}})$ with $U(\varphi_i,b)$.
The Hamiltonian describes two coupled degrees of freedom, whose normal modes form a gapped branch and a gapless branch.
In the next section, we identify the frequency range in which large $C_B$ suppresses corrections to the desired field dynamics.

\section{Breaking flux loops: Decoupling the parasitic mode}\label{app:path_integral}

In Appendix~\ref{app:break}, we introduce additional capacitors to interrupt the large flux loops and reduce the transmission line's sensitivity to magnetic noise. These add a second degree of freedom per unit cell.
Here we use a path-integral formulation to show how the influence of the additional degree of freedom is suppressed for sufficiently large capacitance (small $E_{CB}$) over the dynamical scales of interest.
This modification is particularly suitable for the single-well parameters considered here.
For the double-well parameters, the separation of scales is less pronounced, and the additional mode may have a greater influence on the dynamics.

We consider a $1+1$-dimensional Euclidean field theory with two real fields: $\phi(\tau,x)$, the anharmonic tank-circuit
field, and $\theta(\tau,x)$, the additional circuit field.
The Euclidean Lagrangian density is
\begin{equation}
\begin{split}
  \mathcal{L}_E[\phi,\theta]
  ={}& \frac{1}{2}(\partial_\tau\phi)^2
     + \frac{m^2}{2}\phi^2
     + \frac{\lambda}{4!}\phi^4 \\
     &+ \frac{C_\theta}{2}
        (\partial_\tau\phi-\partial_\tau\theta)^2
     + \frac{1}{2L_\theta}(\partial_x\theta)^2 .
\end{split}
\label{eq:L_start}
\end{equation}
Here $m^2$ is the bare mass-squared parameter and
$\lambda>0$ is the quartic coupling.
We have rescaled both fields by the same factor so that
the tank-circuit kinetic term is canonically normalized.
The dimensionless coefficient $C_\theta=C_B/C$ is the
additional capacitance in units of the tank capacitance,
while $L_\theta^{-1}$ is the spatial-coupling coefficient
in this normalization.

\subsection{Integrating out the additional field}

The action is quadratic in $\theta$, so this field can
be integrated out exactly.
Using the Fourier convention
\begin{equation}
  \phi(\tau,x)
  = \int_{\omega,k} e^{i(\omega\tau+kx)}\phi(\omega,k),
  \qquad
  \int_{\omega,k}\equiv
  \int\frac{d\omega\,dk}{(2\pi)^2},
\end{equation}
and similarly for $\theta$, the quadratic action is
\begin{equation}
  S_{\mathrm{quad}}
  = \frac{1}{2}\int_{\omega,k}
  \begin{pmatrix}\phi^* & \theta^*\end{pmatrix}
  M(\omega,k)
  \begin{pmatrix}\phi \\ \theta\end{pmatrix},
\end{equation}
where
\begin{equation}
  M =
  \begin{pmatrix}
    (1+C_\theta)\omega^2+m^2 & -C_\theta\omega^2 \\
    -C_\theta\omega^2 & C_\theta\omega^2+k^2/L_\theta
  \end{pmatrix}.
  \label{eq:kernel_matrix}
\end{equation}

The Gaussian integral over $\theta$ replaces the quadratic
kernel for $\phi$ by
\begin{equation}
  K_{\mathrm{eff}}
  = M_{\phi\phi}
    - \frac{M_{\phi\theta}^2}{M_{\theta\theta}},
\end{equation}
while leaving the quartic interaction unchanged.
Thus, apart from a field-independent normalization,
\begin{equation}
  S_{\mathrm{eff}}[\phi]
  = \frac{1}{2}\int_{\omega,k}
      K_{\mathrm{eff}}(\omega,k)|\phi(\omega,k)|^2
    + \int d\tau\,dx\,\frac{\lambda}{4!}\phi^4 ,
\end{equation}
with
\begin{align}
  K_{\mathrm{eff}}(\omega,k)
  &= m^2+(1+C_\theta)\omega^2
     -\frac{C_\theta^2\omega^4}
            {C_\theta\omega^2+k^2/L_\theta}
     \nonumber\\
  &= m^2+\omega^2+\frac{k^2}{L_\theta}
     \frac{\omega^2}{\omega^2+\Omega_k^2},
  \label{eq:Keff_basic}
\end{align}
where
\begin{equation}
  \Omega_k^2=\frac{k^2}{C_\theta L_\theta}.
\end{equation}
Here $\Omega_k$ is the characteristic response scale of
the additional field with $\phi$ held fixed; it is not
the exact normal-mode frequency of the coupled system.
The decoupled constant $\theta$ mode is fixed or omitted
from the normalization.
Subject to this convention, the integration over $\theta$
is exact and requires no expansion in $\lambda$.

\subsection{Recovery of the target dynamics}

Equation~\eqref{eq:Keff_basic} can be written as the target
$\phi^4$ kernel plus an explicit correction:
\begin{equation}
  K_{\mathrm{eff}}
  = m^2+\omega^2+\frac{k^2}{L_\theta}
    -\frac{k^2}{L_\theta}
     \frac{\Omega_k^2}{\omega^2+\Omega_k^2}.
\end{equation}
For frequencies and wavenumbers satisfying
\begin{equation}
  r(\omega,k)\equiv
  \frac{\Omega_k^2}{\omega^2}
  =\frac{k^2}{C_\theta L_\theta\omega^2}\ll1,
  \label{eq:decoupling_condition}
\end{equation}
the kernel therefore becomes
\begin{equation}
  K_{\mathrm{eff}}
  = m^2+\omega^2+\frac{k^2}{L_\theta}
    \left[1-r+\mathcal{O}(r^2)\right].
\end{equation}
Within this frequency and wavelength range, the leading
effective action takes the local form
\begin{equation}
\begin{split}
  S_{\mathrm{eff}}[\phi]\simeq
  \int d\tau\,dx\,\bigg\{
    &\frac{1}{2}(\partial_\tau\phi)^2
    +\frac{1}{2L_\theta}(\partial_x\phi)^2 \\
    &+\frac{m^2}{2}\phi^2
    +\frac{\lambda}{4!}\phi^4
  \bigg\}.
\end{split}
\end{equation}
This is the desired local $\phi^4$ action.
At fixed nonzero $\omega$ and fixed $k$, the leading
correction decreases as $1/C_\theta$.
Physically, a large capacitance suppresses relative
time-dependent motion of the two fields, allowing $\phi$
to inherit the spatial coupling of $\theta$ over this range.

The relevance of this separation to the single-well regime
can be illustrated in the quadratic theory, with
$\lambda=0$ and $m^2>0$.
After continuation to real frequency $\nu$, the normal-mode
frequencies satisfy
\begin{equation}
  \nu^4
  -\left[
      m^2+\left(1+\frac{1}{C_\theta}\right)
      \frac{k^2}{L_\theta}
    \right]\nu^2
  +m^2\Omega_k^2=0.
\end{equation}
The upper branch obeys
$\nu_+^2(k)\geq m^2+k^2/L_\theta$, and hence
$\Omega_k^2/\nu_+^2(k)\leq1/C_\theta$.
For $C_\theta\gg1$, its dispersion therefore approaches
that of the target quadratic theory.
For interacting dynamics, Eq.~\eqref{eq:decoupling_condition}
specifies the range in which the quadratic part of the
exact effective action is approximated by the target kernel.

This approximation is not uniform in the static limit.
At $\omega=0$ and $k\ne0$, the exact kernel instead gives
$K_{\mathrm{eff}}(0,k)=m^2$: capacitive coupling constrains
relative motion, but does not lock static field profiles.
The result therefore establishes suppression over the
specified dynamical scales, rather than decoupling at all
frequencies.
Slow dynamics, including those relevant to double-well
configurations, need not lie within this regime.
\bibliography{refs}

\end{document}